# Propuesta de abordaje transversal del concepto diferencial de curvatura en situaciones físicas

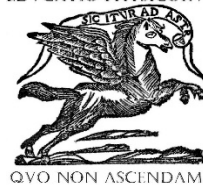

**Mauricio López Reyes[1,2]**
[1]*Departmento de investigación, Instituto Frontera. Calle Constitución No. 10,000. Colonia Centro, C.P. 22,000, Tijuana, B. C. México.*
[2]*Departamente de Física de la Tierra y Astrofísica. Facultad de Ciencias Físicas. Universidad Complutense de Madrid. Av. Séneca, 2, 28040 Madrid, España.*

**E-mail:** maurilop@ucm.es

**Resumen**

Durante el proceso de enseñanza del concepto de derivada, es común y natural hacer referencia a interpretaciones geométricas, tal es el caso de la utilización de la recta tangente y los puntos máximos y mínimos de una función, para ilustrar los alcances de la derivada. En este trabajo, se presenta una manera incluyente y aplicada del concepto de diferencial de curvatura, específicamente con la intervención de problemas de la mecánica de Newton, cuyo propósito es doble, por un lado, dotar a la derivada de un sentido altamente versátil para modelar sistemas, y por otro; abordar diversos temas clásicos de los cursos de física inicial desde una propuesta didáctica centrada en el concepto de curvatura.

**Palabras clave:** Enseñanza de la mecánica; Curvatura de una función; Apicaciones física de la derivada; Modelación de sistemas.

**Abstract**

During the process of teaching the concept of derivative, it is common and natural to refer to geometric interpretations, such as the use of the tangent line and the maximum and minimum points of a function, to illustrate the scope of the derivative. In this work, an inclusive and applied way of the concept of curvature differential is presented, specifically with the intervention of problems of Newton's mechanics, whose purpose is double, on the one hand, to provide the derivative with a highly versatile sense to model systems, and on the other hand; to approach diverse classic topics of the initial physics courses from a didactic proposal centered on the concept of curvature.

## I. INTRODUCCIÓN

En la realidad de los estudiantes de los primeros semestres de universidad de carreras científicas o de ingeniería, una de las principales preocupaciones es el aprender a derivar e integrar, y con cierto éxito esto se cumple, ya que los alumnos muestras sus habilidades para aplicar las diversas "fórmulas" de derivación e integración, lo anterior debido a que se cuenta con la premisa de que estas matemáticas serán aplicadas en sus cursos de física posteriores o incluso contemporáneos. Sin embargo, esta misión de aprender a derivar o integrar, se vuelve un problema intrínseco en la enseñanza del cálculo, simplificado a sólo aplicar fórmulas; cuando la inspiración de Newton y Leibniz fue crear una herramienta capaz de describir la naturaleza. Si bien es cierto que hay diferentes interpretaciones de la derivada, por ejemplo, las geométricas, como la pendiente de una función en un punto o problemas de razones de cambio instantáneas, la realidad física de nuestro universo, al menos la mecánica, es descrita de forma muy simple y elegante con estructuras que implican a la derivada vista como un operador y como un agente que mide el cambio.

Acerca de la importancia del cálculo y su relación con la enseñanza de la física, [1] expone:

*Si el Cálculo diferencial ha resultado imprescindible para el desarrollo científico, y en particular para el avance en la Física, es lógico que también resulte imprescindible, a partir de ciertos niveles, en la enseñanza de la Física, cuando hayan de tratarse situaciones mínimamente complejas, más cercanas a la realidad que las tratadas en cursos elementales. El inicio en el uso del Cálculo diferencial en la enseñaza de la Física se produce en el último año de Bachillerato donde más del 80% de una muestra de 103 profesores de Física reconocen abiertamente que necesitan el Cálculo diferencial para desarrollar la asignatura. Ya en el nivel universitario, el Cálculo está presente en la práctica totalidad de los tópicos de Física.*

Históricamente, el desarrollo del cálculo se ha basado en las demandas de la física de contar con nuevas herramientas para predecir la evolución de los fenómenos naturales [2] pero al momento de tomar los cursos de cálculo en las Instituciones de nivel superior e incluso de nivel medio superior, la *praxis* es distinta en muchos de los casos, como se mencionó previamente, el docente se centra en la derivada como un operador con algunas implicaciones geométricas, pero muchas veces desligado de su representación física. Tal vez, uno de los mayores retos de las instituciones de nivel superior en donde se ofertan carreras de ciencias básicas e ingenierías es que la docencia de las asignaturas de matemáticas corren a cargo de personal docente con una alta preparación en matemáticas teóricas, es decir, son cursos con

*Mauricio López Reyes*

bastante rigor [3] esto no es un problema en lo absoluto, de la misma manera, los cursos de física se imparten bajo los docentes con una alta especialidad en física, incluso con trayectoria en investigación, la situación retadora versa en que tanto el profesor de matemáticas como el de física presentan sus contenidos temáticos muchas veces desvinculados a la realidad de madurez académica del estudiante, y las frases como *"La demostración queda para el alumno"* o *"Solamente es expandir en series de Taylor y resolver para la energía"* son cotidianas, además de que las actividades de aprendizaje en los cursos de física presuponen una conexión obvia entre las matemáticas y la descripción del fenómeno y esto no necesariamente evidente para el alumno.

Es por ello que, el objetivo de este trabajo es presentar una propuesta didáctica para estrechar el vínculo entre un concepto netamente geométrico como lo es la curvatura, pasando por su análisis diferencial y sus aplicaciones en problemas de la mecánica de Newton para lograr el objetivo de mostrarle al estudiante que los conceptos matemáticos y físicos que aprende en sus primeros cursos, cobran mayor fuerza cuando se abordan de manera cohesionada y esto detone mayor interés y significancia en los aprendizajes esperados.

Una de las aplicaciones geométricas que comúnmente se abandonan en un curso de cálculo diferencial, es el problema de la curvatura, ya que implica un dominio no solo operacional de la derivada, sino un nivel de comprensión conceptual y geométrico relativamente avanzado para los alumnos de primer curso. Dicho concepto geométrico de curvatura tiene múltiples aplicaciones en diversas áreas de la física. Se precisa que, en mecánica clásica el temario abarca temas de cinemática, estática, dinámica lineal y dinámica circular, donde el concepto de derivada aparece numerosas cantidades de veces, especialmente al referirse a los conceptos de velocidad y aceleración, de esta manera, el concepto geométrico de curvatura embona perfectamente en los temas de cinemática y dinámica curvilínea o circular.

Por otro lado, la Teoría de los Sistemas Complejos formulada por García, enfatiza la noción de interdisciplina como proceso. En este trabajo el concepto de interdisciplina se logra al enlazar los conceptos geométricos y diferenciales de la curvatura, con casos aplicados en la física lo cual para [3,4] lo convierte en un sistema complejo.

Al revisar los planes de estudio tanto de cálculo diferencial como de mecánica clásica, es pertinente mencionar que los temas de cinemática y dinámica curvilínea puede ser abordados en la segunda mitad del curso, sin sacrificar perdida de ilación de temas; esto brindaría mayor tiempo para lograr niveles superiores de aprendizaje según la taxonomía de Bloom, y su transferencia al concepto aplicado de derivada, además para que sus implicaciones físicas sean más naturales [5], lo anterior brinda la oportunidad de que, el concepto de curvatura aplicado a situaciones de la mecánica de Newton detone los aprendizajes significativos esperados para el estudiante [6].

Los casos de aplicación física que se expondrán como transferencia del concepto de curvatura son: A) movimiento curvilíneo uniforme (MCU) y su la dinámica rotacional y B) movimiento curvilíneo sobre una trayectoria parabólica; ambas apoyadas por sus respectivas deducciones geométricas y modelación gráfica mediante en lenguaje de Python.

Todos los problemas físicos mencionados anteriormente serán abordados desde el punto de vista de la geometría plana, i.e., la definición que se construirá de curvatura es válida en el plano euclideo y con sistemas de coordenadas cartesianas, no obstante su generalización a un sistema de más dimensiones o con otros sistemas de coordenadas no es necesaria para el nivel de complejidad de los problemas físicos que se abordan en los primeros cursos de física universitaria, para los cuales se piensa que será útil esta propuesta.

## II. CONCEPTO GEOMÉTRICO Y DIFERENCIAL DE CURVATURA

En esta sección abordaremos la deducción de la definición geométrica y diferencial de la curvatura, así como de su parámetro implícito conocido como radio de curvatura. Esta parte netamente matemática es la correspondiente al curso de cálculo diferencial, típicamente se ubica en los libros de texto tradicionales en los últimos capítulos, aún después de "máximos y mínimos". El objetivo es ser muy minucioso en el desarrollo de la función de curvatura para después hacer la transferencia a las implicaciones físicas. Tal y como [5] lo mencionan, este trabajo pretende redescubrir las formas de enseñanza de conceptos físicos a través de situaciones problemáticas, en este caso bajo la de la necesidad de definir objetos matemáticos como lo es la curvatura.

### A. Construcción de la función de curvatura en el plano

Consideremos una función $f(x)$, continua y suave en el intervalo comprendido entre $[a, b]$, donde $a, b \in \mathbb{R}$.

En este punto, vale la pena mencionar que la implicación física de la continuidad representa que el fenómeno que estamos estudiando evoluciona de manera continua a través del tiempo y/o del espacio, es decir, no hay huecos en el tiempo ni en el espacio. Para el caso de la demanda de que la función sea suave, su implicación física es que no tendremos colisiones elásticas y por lo tanto, la taza de cambio de su posición (por ejemplo) presenta cambios graduales.

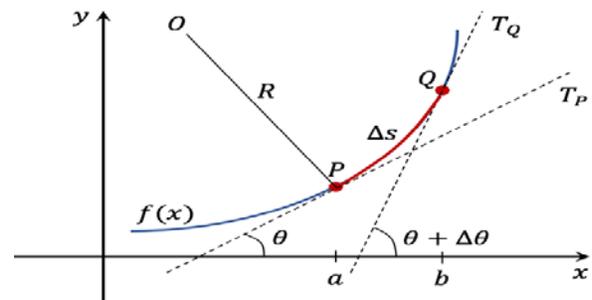

**FIGURA 1.** Función $f(x)$, continua y suave en el intervalo $[a, b]$. También se observan dos puntos $P$ y $Q$, en donde cortan dos rectas tangentes $T_P$ y $T_Q$, y los ángulos que forman con el eje horizontal $\theta$

y $\theta + \Delta\theta$, respectivamente, y la longitud de arco $\Delta s$, comprendida entre los puntos.

## II. ILUSTRACIONES

En la figura 1, se observa la función $f(x)$, así como los puntos $P$ y $Q$, por donde cortan rectas tangentes a cada uno de ellos $T_P$ y $T_Q$, respectivamente. Además, debido al incremento en la pendiente de la recta $T_Q$, respecto a $T_P$, el ángulo $\theta$, de $T_P$, se incremente a $\theta + \Delta\theta$, de $T_P$, por lo que se puede inferir que la función experimenta un cambio gradual en su pendiente entre los puntos $P$ y $Q$, justo a este cambio se le define como curvatura de la función y puede advertirse que entre mayor sea el cambio en los ángulos, mayor será la curvatura.

De manera formal, definimos la curvatura de una función en $\mathbb{R}^2$ como sigue:

**Definición 1**. *La curvatura $\kappa$, de una función $f: \mathbb{R} \to \mathbb{R}$, definida por $y = f(x)$, en el punto $P$, se define como el cambio instantáneo de un incremento del ángulo de las rectas tangentes a los puntos $P$ y $Q$, respecto al incremento de la longitud arco $\Delta s$.*

$$\kappa = \lim_{\Delta s \to 0} \frac{\Delta \theta}{\Delta s} = \frac{d\theta}{ds}. \quad (1)$$

En este momento es interesante aclarar que, de no definir la curvatura como el límite anterior, no significa que no sea la curvatura, sino que la curvatura calculada sería su valor medio entre los puntos $P$ y $Q$, cuestión que pasaría desapersivida por ejemplo, en la circunferencia, donde la razón de cambio media es igual que la instantánea.

Al tomar la parte diferencial de la expresión (1), y utilizando la regla de la cadena, se logra lo siguiente,

$$\kappa = \frac{d\theta}{ds} = \frac{d\theta}{dx} \cdot \frac{dx}{ds} = \frac{d\theta}{dx} \cdot \frac{1}{\frac{ds}{dx}}, \quad (2)$$

además, de la geometría de la figura 1, tenemos por la definición de tangente que $dy/dx = \tan\theta$, por lo tanto,

$$\theta = \arctan\left(\frac{dy}{dx}\right), \quad (3)$$

en (3) tenemos una relación explícita de $\theta$, en función de la derivada de la función, por lo que al volverla a derivar respecto a $x$, obtenemos la razón de cambio instantánea entre estas dos variables.

$$\frac{d\theta}{dx} = \frac{1}{1 + \left(\frac{dy}{dx}\right)^2} \cdot \frac{d^2 y}{dx^2}. \quad (4)$$

Con ayuda de la figura 2, y mediante el teorema de Pitágoras, es posible observar que la derivada de la longitud de arco respecto a $x$ es la expresión (5).

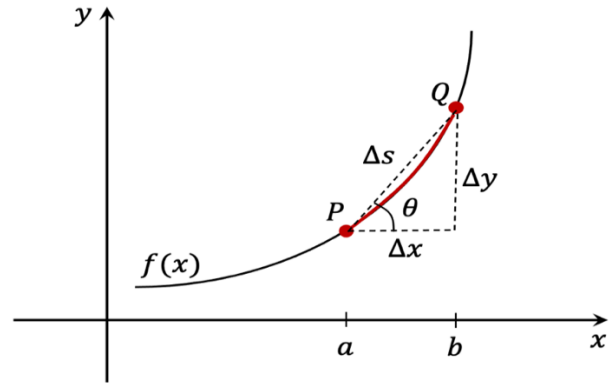

$$\frac{ds}{dx} = \sqrt{1 + \left(\frac{dy}{dx}\right)^2}$$

**FIGURA 2.** Función $f(x)$ y elemento diferencial de longitud de arco $\Delta s$.

Al sustituir (4) y (5) en (2), tenemos

$$\kappa = \frac{\frac{d^2 y}{dx^2}}{1 + \left(\frac{dy}{dx}\right)^2} \cdot \frac{1}{\sqrt{1 + \left(\frac{dy}{dx}\right)^2}}, \quad (5)$$

al reducir la ecuación anterior, obtenemos la expresión diferencial para la curvatura de una función en el punto $P$.

$$\kappa = \frac{\frac{d^2 y}{dx^2}}{\left[1 + \left(\frac{dy}{dx}\right)^2\right]^{3/2}}. \quad (6)$$

Adviértase que (6) se puede comportar como una función de curvatura que depende del valor $x$, la notación tradicional es mostrar a $\kappa$ en función de $x$, como en la expresión (7),

$$\kappa(x) = \frac{\frac{d^2 y}{dx^2}}{\left[1 + \left(\frac{dy}{dx}\right)^2\right]^{\frac{3}{2}}}, \quad (7)$$

de (7) se deduce que $\kappa > 0$, cuando el punto $P$, está sobre un arco cóncavo, y $\kappa < 0$, cuando pertenece a un arco convexo, sin embargo, en la aplicación de los problemas mecánicos que analizaremos, el signo no tiene relevancia.

**Definición 2**. *El radio de curvatura $R$, de una función continua y suave en el punto $x$, es inversamente proporcional a su curvatura.*

$$R = \frac{1}{\kappa}, \quad \kappa \neq 0. \quad (8)$$

*Mauricio López Reyes*

Este elemento se puede observar en la figura 1, con la longitud de segmento $\overline{OP}$, donde el punto $O$, se conoce como el centro del círculo osculador. De la ecuación (8) es importante mencionar una consideración geométrica algo oculta, y es el hecho de que, si la curvatura de una función es cero, tal es el caso de la línea recta, entonces su radio de curvatura será infinito, es decir, si tomamos el límite siguiente $\lim_{\kappa \to 0} 1/\kappa = \infty$, dicho resultado tiene mucho sentido, ya que el círculo osculador será infinitamente grande lo cual se traduce en una curva donde el ángulo de la pendiente es constante. Justo este tipo de detalles matemáticos son lo que cobran gran interés en la aplicación de problemas físicos; de la misma manera, usted podría preguntarse lo siguiente ¿cuál será el punto de mayor curvatura?, y ¿qué significado físico tiene dicho punto? Tal y como se enseña en cálculo diferencial, este es un problema de optimización, y se tiene que tomar la derivada de la función (7) e igualarla a cero para encontrar el punto de máxima curvatura. Respecto a su interés físico, lo veremos con detalle en los ejemplos de aplicación propuestos.

Con base en la expresión (8), bien la podríamos pensar en sentido recíproco, i.e., que la función de curvatura depende del inverso del radio, $\kappa = 1/R$, esta expresión nos brindará mayor soporte para analizar algunos casos de la dinámica de Newton.

## III. PROPUESTA DE APLICACIÓN EN PROBLEMAS DE MECÁNICA

Una vez deducido el concepto de curvatura, como típicamente se hace en un curso de cálculo diferencial, sería trabajo del docente de física introducir este concepto dentro de sus secuencias didácticas; se recomienda utilizar los temas relacionados con la cinemática y dinámica rotacional como motivadores para inducir al estudiante en situaciones problemáticas que le permitan llegar a los niveles más altos de aprendizaje, por ejemplo, en la taxonomía de Bloom sería analizar, evaluar y crear [7]. Específicamente para los fines de esta propuesta, se pretende que el profesor, conduzca al alumno a analizar la situación física propuesta y su relación con el concepto de curvatura formulando una propuesta de modelo algebraico para dar solución al problema, evaluar la congruencia y pertinencia del concepto de curvatura como herramienta para modelar los sistemas, y crear un modelo teórico que permita representar la realidad física del fenómeno a través de un concepto geométrico como el de curvatura.

En los siguientes ejemplos se expone una alternativa para estudiar algunos problemas de mecánica abordados desde el punto de vista de la curvatura, dentro del desarrollo de los ejemplos se proponen las siguientes secciones: i) planteamiento de la situación física, ii) introducción de la curvatura en la situación física, iii) análisis general del modelo incluyendo la curvatura y estudio de los casos particulares y iv) comentarios y recomendaciones didácticas.

### A. movimiento curvilíneo uniforme (MCU) y dinámica rotacional

i) Consideremos una partícula de masa $m$, que se mueve con rapidez constante a lo largo de una trayectoria circular, cuyo radio es $R$, como se observa en la figura 3.

Con base en la definición de aceleración centrípeta, tenemos que,

$$a_c = \frac{v^2}{R}. \qquad (9)$$

Donde $v$, es la magnitud de la velocidad y $R$, el radio de la trayectoria circular, i.e., el radio de curvatura.

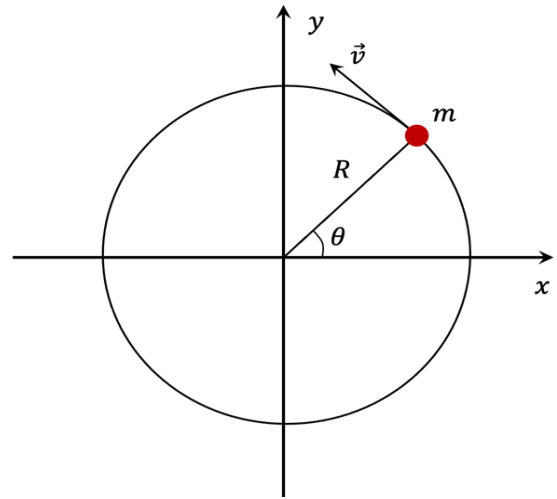

**FIGURA 3.** Partícula moviéndose sobre una circunferencia de radio inicial $R$.

ii) Por lo anterior, podemos sustituir (8) en (9) para obtener la siguiente expresión para la aceleración centrípeta en función de la curvatura:

$$a_c = \kappa v^2, \qquad (10)$$

en (10), se observa que la aceleración centrípeta es proporcional linealmente a la curvatura de la circunferencia, esto lo podemos escribir en forma de una función de forma explícita, al hacer $a_c(\kappa) = \kappa v^2$.

iii) Con base en (10), es intuitivo pensar que la gráfica de la aceleración centrípeta en función de la curvatura será una línea recta con pendiente igual a $v^2$, tal y como se observa en la figura 4(a). Si al mismo tiempo la rapidez es variable; entonces tendremos una función de 2 variables independientes $a_c(\kappa, v)$, cuyo modelo se presenta en la figura 4(b), se observa como tenemos dependencia lineal y cuadrática respecto a la curvatura y rapidez respectivamente.

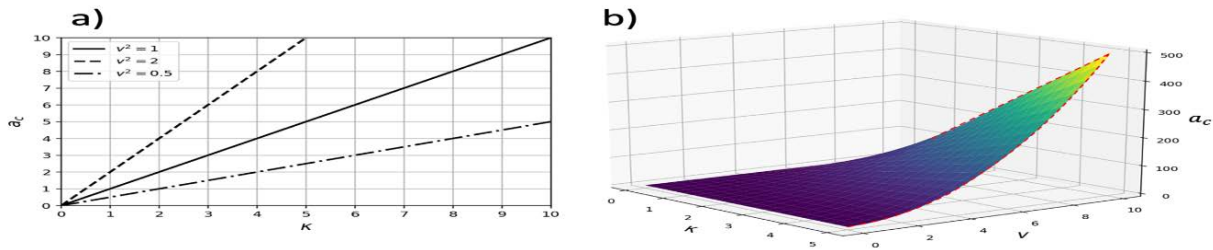

**FIGURA 4.** (a) Dependencia lineal de la $a_c$, respecto a la curvatura de la trayectoria circular. Se presentan 3 gráficas para distintos valores de la rapidez. (b) Modelación de $a_c(\kappa, v)$, se observa el comportamiento lineal en función de $\kappa$ y $v$, respectivamente.

Como parte del análisis transversal entre el cálculo diferencial y la parte física, vale la pena derivar la expresión (10) respecto a la curvatura, por lo que obtenemos (11),

$$\frac{\partial a_c}{\partial \kappa} = v^2. \quad (11)$$

Esta sencilla operación desde el punto de vista del cálculo nos permite extraer aún más información física del sistema, y es lo que muchas veces en la práctica docente se omite por diversas situaciones, entre ellos destaca el dar por sabido o trivial dicha implicación, lo cual es reafirmado por [8].

La oportunidad que el profesor de física puede aprovechar en este momento es muy importante para reforzar el significado de la derivada, podría por ejemplo, preguntar a la clase lo que significa geométricamente la derivada, esperando que los alumnos la identifiquen como la pendiente de la recta tangente en un punto; entonces la transversalidad viene en la pregunta ¿qué significa físicamente la derivada $\partial a_c/\partial \kappa$?, al transferir la definición de derivada desde su definición geométrica hacia su interpretación física, sólo se tendría que hacer una sustitución de palabras, para decir que la derivada (11) representa la razón de cambio (pendiente) de la aceleración centrípeta respecto a la curvatura, de aquí se infieren 2 situaciones particulares. 1) si $v \rightarrow 0$, implica que $\partial a_c/\partial \kappa \rightarrow 0$, i.e., si la rapidez lineal de la partícula tiende a cero, entonces la taza de variación de $a_c$, respecto a $\kappa$ también lo hará, por lo que no importa la curvatura que tenga la circunferencia, la aceleración centrípeta no cambiará. 2) si $v \gg 1$, implica que $\partial a_c/\partial \kappa \gg 1$, es decir, la razón de cambio de (11) aumenta cuadráticamente en función de $v$, lo cual tiene total sentido con su definición.

En este mismo análisis transversal del concepto de aceleración centrípeta y curvatura, es interesante incorporar saberes previos del alumno, tal es el caso de la relación algebraica entre rapidez lineal y angular, deducida a partir de la longitud de arco,

$$v = \omega R, \quad (12)$$

donde $\omega$, es la rapidez angular de la partícula. Al sustituir (12) y (8) en (9), obtenemos una función de aceleración centrípeta totalmente equivalente a (10).

El análisis realizado para (10) lo podemos repetir con (13) y obtendríamos resultados equivalentes; pero ahora en función de la velocidad angular. Para este caso, las gráficas de la función (13) se presentan en la figura 5(a) y (b) para el modelaje respecto a $\kappa$, con $\omega = cte$, y con $\kappa$ y $\omega$, variables respectivamente.

$$a_c(\kappa, \omega) = \frac{\omega^2}{\kappa}. \quad (13)$$

iv) El desarrollo de este ejemplo deja de manifiesto la estrecha, pero pocas veces mencionada relación entre la curvatura y el concepto físico de la aceleración centrípeta en mecánica clásica. Además, se muestra que, con sólo aplicar algunas cuentas "sencillas" como lo son las derivadas de funciones polinomiales, es posible inferir consecuencias físicas a partir del concepto de curvatura. Se recomienda que, en la práctica docente se motive al alumno a abordar un problema físico de forma creativa y no limitarse a la pura tradición formal de los libros básicos, especialmente cuando el aprendizaje esperado involucra las escalas más altas de la taxonomía de Bloom, como se esperaría que un estudiante de educación superior cumpla satisfactoriamente. De la misma manera, el complemento de la modelación gráfica brinda una ventaja para asimilar conceptos que podrían ser abstractos, como la interpretación de la derivada en un contexto fuera de lo geométrico.

Otro de los grandes retos en la enseñanza de la física universitaria y de niveles educativos inferiores, es sin duda, la ilación con la que se presentan los temas, ya que muchas veces, la exposición docente esta definida a trozos o de manera discontinua, aunque el temario se cumpla por completo. Siguiendo el tema de curvatura y aceleración centrípeta, es intuitivo extender la transversalidad a un tema propio de la dinámica como las leyes de movimiento de Newton, especialmente con el concepto de fuerza centrípeta que, en el caso más simple se podría definir como *la fuerza que experimenta una partícula con rapidez constante, pero con la dirección de la velocidad variable*, lo que implica trayectorias curvas con cierto radio de curvatura, y justo aquí es donde entra en acción la conexión entre la cinemática y la dinámica curvilínea, y la curvatura es el eslabón que hilará las ideas principales.

(13)

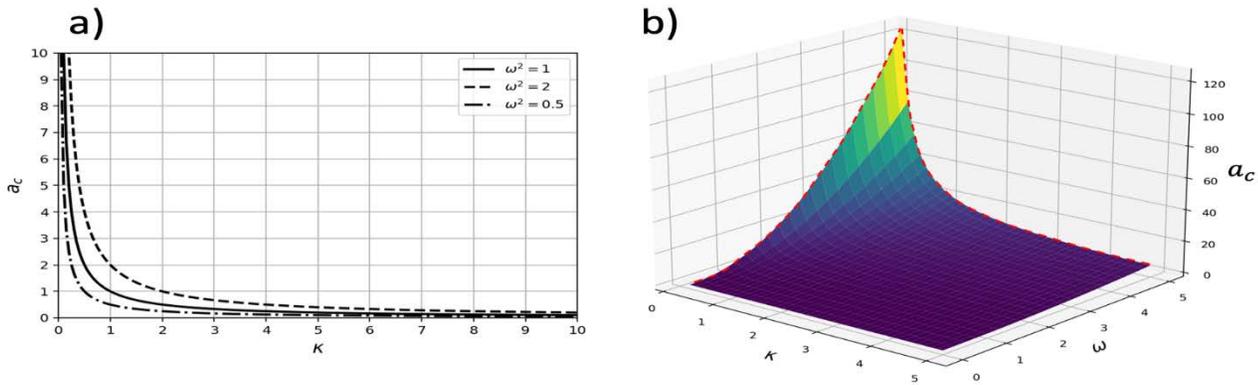

**FIGURA 5.** (a) Dependencia hiperbólica de la $a_c$, respecto a la curvatura de la trayectoria circular. Se presentan 3 gráficas para distintos valores de la rapidez angular. (b) Modelación de $a_c(\kappa, \omega)$, se observa el comportamiento hiperbólico y parabólico respecto de $\kappa$ y $\omega$, respectivamente.

***Definición 3.*** *La fuerza centrípeta se define como la componente de la fuerza debida al movimiento curvilíneo y se dirige hacia el centro de curvatura.*

Algebraicamente su expresión escalar nace de la segunda ley de Newton para el caso curvilíneo, tal y como se presenta en (14).

$$F_c = m a_c. \qquad (14)$$

Al sustituir las expresiones de la cinemática curvilínea, y la definición de curvatura en función del radio, específicamente las ecuaciones (8) y (9), tenemos la expresión para la magnitud de la fuerza centrípeta en función de la rapidez y la curvatura en (15),

$$F_c = m \kappa v^2 \qquad (15)$$

donde $F_c$, es la magnitud de la fuerza centrípeta y $m$, la masa de la partícula. Por comparación con (10), se observa que la magnitud de la fuerza depende en la misma proporción de la curvatura y la rapidez que la aceleración centrípeta, por lo que el modelaje gráfico es equivalente al presentado en la figura 4, la única diferencia es el término de la masa. Es claro de (15) que, al aumentar la curvatura de la trayectoria (seguimos pensando que es circular) entonces la fuerza centrípeta aumenta linealmente; se insinúa al docente que pregunte a los alumnos sobre que variable influye más en la fuerza centrípeta, la masa, la curvatura o la rapidez, este ejercicio le brindará al alumno la oportunidad de pensar sobre los órdenes de magnitud relacionados con las cantidades "típicas" macroscópicas que se manejan en mecánica clásica, además; el docente podría proponer a los alumnos la realización de las derivadas parciales (16) y su interpretación física.

$$\frac{\partial F_c}{\partial m} = \kappa v^2, \quad \frac{\partial F_c}{\partial \kappa} = m v^2 \quad y \quad \frac{\partial F_c}{\partial v} = 2 m \kappa v. \qquad (16)$$

Por experiencia docente, es recomendable recalcar en clase que la derivada nos informa directamente como cambia la fuerza centrípeta en relación con cada una de las variables, por lo que la respuesta a la pregunta planteada previamente sobre *¿qué influye más en la fuerza centrípeta?* se puede abordar desde el punto de vista de las derivadas (16), aquella que tenga una pendiente mayor (o simplemente mayor valor) es la que influye más en el cambio de la fuerza centrípeta. Advertencia, estos análisis deben tomarse con precaución ya que podrían no ser válidos en todo el dominio de la función (15).

Por ejemplo, al observar $\partial F_c / \partial \kappa$ y $\partial F_c / m$, es directo que ambas derivadas son "semejantes" en el sentido de que ambas tienen una variable lineal por la rapidez al cuadrado, por lo que si $\kappa > m$, entonces $\partial F_c / \partial \kappa > \partial F_c / m$, y biseversa; este tipo de análisis pondrá a prueba al alumno, ya que tendrá que hacer una doble comparación, una netamente matemática al recordar y aplicar el concepto de derivada, y la otra buscando la interpretación física de cada variable. Sólo por ir un poco más allá, si el curso de física se toma simultaneo con cálculo de vectorial, una situación problemática interesante y de mucho provecho práctico es preguntar al alumno ¿qué significado tendría el gradiente de la fuerza centrípeta? e incluso calcular su magnitud y campo de direcciones. Hasta antes de esta pregunta, ya se tienen todos los ingredientes para aplicar el gradiente, sólo haría falta introducir algo de notación vectorial, para llevar las expresiones de (16) a (17),

$$\nabla F_c = \frac{\partial F_c}{\partial m} \hat{m} + \frac{\partial F_c}{\partial \kappa} \hat{\kappa} + \frac{\partial F_c}{\partial v} \hat{v}, \qquad (17)$$

como acabamos de mostrar, las implicaciones que tiene el concepto de curvatura en un problema elemental de mecánica pueden proporcionar material para varias secuencias didácticas durante el curso, siempre con el objetivo de motivar al alumno a que aprender matemáticas le brindará de una valiosa herramienta para entender, analizar, aplicar, evaluar y crear conocimiento.

**B. Movimiento curvilíneo sobre una trayectoria parabólica**

Consideremos una partícula de masa $m$, que se mueve sobre una trayectoria descrita por la función $y(x) = C x^2$, donde $C$, es una constante, para el desarrollo del ejemplo vamos a considerar que la rapidez con la que se mueve la partícula es constante e igual a $v$. En la figura 6, se presentan diferentes trayectorias dependientes del valor de la constante arbitraria $C$.

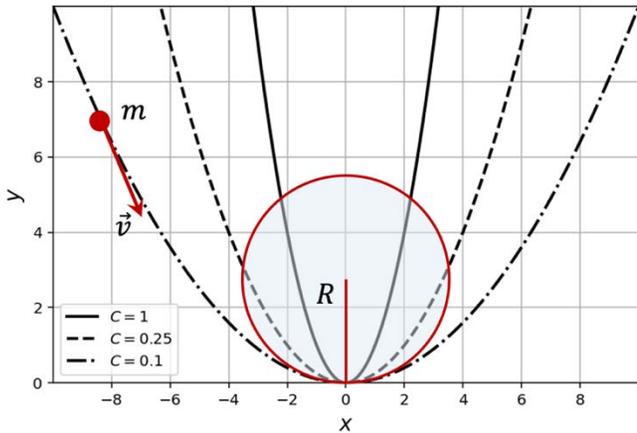

**FIGURA 6.** Partícula moviéndose sobre una parábola $y(x) = Cx^2$, para distintos valores de $C$. Se presenta un ejemplo de la partícula de masa $m$, sobre la trayectoria con $C = 0.1$ y cuyo radio de curvatura y círculo osculador se presenta cuando $x = 0$.

i) La situación física que motiva el problema es determinar una expresión para la fuerza centrípeta que experimenta la partícula en las trayectorias presentadas en función de su curvatura, desde luego modelando gráficamente su comportamiento.

Partimos de la definición de fuerza centrípeta aplicada en (15), en donde se presenta como función de la curvatura, masa y rapidez que estamos asumiendo constante.

$$F_c = m\kappa v^2.$$

ii) Para este caso, en que tenemos una partícula moviéndose en una trayectoria parabólica descrita por $y(x)$, podemos calcular la función de curvatura aplicando su definición expuesta en (7). Al efectuar las operaciones indicadas y simplificar la expresión, llegamos a la función de curvatura (18).

$$\kappa(x) = \frac{2C}{[1 + 4C^2 x^2]^{\frac{3}{2}}}. \quad (18)$$

Nótese que $\kappa(x)$, también depende del valor de la constante, por lo que la notación correcta sería $\kappa(x, C)$. Es claro que si $C \to 0$, la curvatura disminuye, ya que $R \to \infty$, lo que se interpreta como una parábola muy "achatada" pareciéndose a una línea recta, cuya curvatura es nula; de manera contraria, si $x \to 0$, entonces la $\kappa$, será máxima, esto se puede demostrar derivando la función de curvatura (18) igualarla a cero y resolver para $x$, tal y como se hace en cálculo diferencial en el tema optimización, dicho ejercicio sería de gran valor para los estudiantes. No es de sorprender que el punto de mayor curvatura es en $x = 0$, independientemente de $C$, ya que en ese punto el radio de curvatura es mínimo. Sin embargo la constante $C$, influye en cada caso particular pero en todos ellos el punto de máxima curvatura será el mismo.

Este análisis condensado que se acaba de exponer, presupone algunas cuentas que el docente o el alumno debe realizar para demostrar lo dicho.

iii) Ahora, retomamos la situación física, sustituyendo (18) en la expresión para la fuerza centrípeta dada en (15), con lo cual obtenemos la función de fuerza centrípeta dependiente explícitamente de $x$, tal y como se observa en (19).

$$F_c(x) = \frac{2Cmv^2}{[1 + 4C^2 x^2]^{\frac{3}{2}}}. \quad (19)$$

De (19), se pueden hacer los siguientes análisis de casos límite:

1. Si $x \to 0$, entonces $F_c \to 2Cmv^2$,
2. Si $x \to \infty$, entonces $F_c \to 0$.

Lo que tiene mucho sentido intuitivo ya que el término $4C^2 x^2$, es el que cambia en función de $x$, y por lo tanto si aplicamos los casos límite 1 y 2, llegamos a las conclusiones realizadas sobre la curvatura al inicio de ii).

Tal y como se mencionó en el análisis de la curvatura, para determinar el punto de máxima fuerza centrípeta, derivamos (19) respecto de $x$, igualamos a cero y resolvemos para la misma variable,

$$\frac{\partial F_c}{\partial x} = \frac{-2Cmv^2 \left[\frac{3}{2}(1 + 4C^2 x^2)^{1/2} \cdot 8C^2 x\right]}{[1 + 4C^2 x^2]^3} = 0,$$

$$-24C^3 mv^2 x \sqrt{1 + 4C^2 x^2} = 0, \quad (20)$$

al resolver la ecuación para $x$, obtenemos 2 posibles soluciones, $x_1 = 0$, y $x_2 = -i/2C$, que naturalmente la que tiene significado físico es $x_1 = 0$, tal y como se había planteado con anterioridad. Por lo tanto, la fuerza centrípeta será máxima cuando $x = 0$, y la magnitud correspondiente de la fuerza centrípeta máxima estará dara por la expresión (21).

$$F_{c,máx} = 2Cmv^2. \quad (21)$$

Note que (21) es justo lo que esperamos, debido a que si $C$, aumenta, también lo hace la curvatura y por lo tanto la fuerza centrípeta también lo hará. Con el objetivo de modelar algunos casos particulares y mostrar la veracidad gráfica de lo que se acaba de demostrar, vamos a plantear 3 situaciones que se presentan a continuación:

En la situación (a) vamos a hacer: $m = 1.0\ kg, v = 1.0\ ms^{-1}$ y $C = 1\ m^{-1}$, en las situaciones (b) y (c) mantenemos constante la masa, pero $v = 2.0\ ms^{-1}$ y $C = 0.5\ m^{-1}$ y $v = 4.0\ ms^{-1}$ y $C = 0.25\ m^{-1}$ respectivamente. El modelaje de cada una de las situaciones anteriores se presenta en la figura 7.

En este punto es natural que el alumno o incluso el profesor se pregunte ¿cómo afectan y en qué proporción las 4 variables de la figura 7?, para ello es recomendable realizar las gráficas de la fuerza centrípeta en función de cada una de ellas o incluso se podría hacer un modelaje tridimensional en combinación de pares de variables independientes, esto dependerá en gran medida del nivel matemático del estudiante, los softwares para graficación en tres dimensiones y/o de las herramientas de programación conocidas.

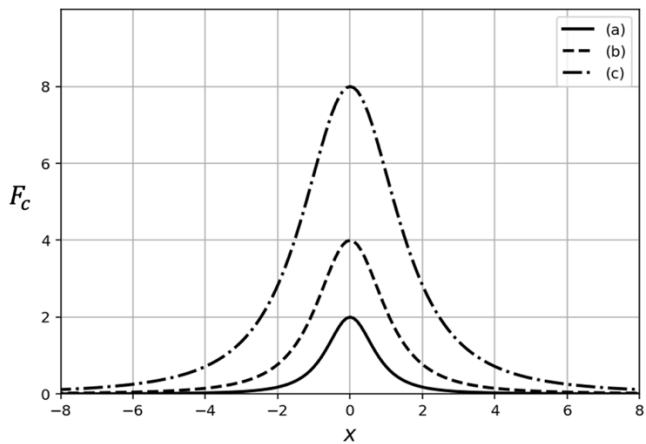

**FIGURA 7.** Comportamiento de la fuerza centrípeta en función de las variables $m, v, C$ y $x$. Implícitamente está involucrada la curvatura.

Por ejemplo, en la figura 8, se presenta el comportamiento de la fuerza centrípeta en función de la posición $x$ y el valor de $C$, ambos parámetros influyen en el valor de la $\kappa$, tal y como vimos en (18), pero en la figura (8) se descompone la influencia de cada uno de ellos. Si hacemos cortes paralelos al eje $x$, se observa que las diferentes intersecciones para los valores de $C$, coinciden con la gráfica de la figura 7. Al comparar la variación respecto a $x$ y $C$, se espera que el alumno concluya que los máximos valores de la fuerza centrípeta se alcanzan sobre $x = 0$ y a medida que $C$ crece, adviértase como en esta situación de análisis vuelve a aparecer el concepto de gradiente como la dirección de máximo crecimiento de la fuerza centrípeta.

iv) Finalmente, esta propuesta didáctica y de transversalidad puede enriquecer potencialmente el proceso de enseñanza-aprendizaje si además de brindarle todos estos conocimientos al alumno, se fomenta su pensamiento analítico con preguntas como las siguientes: ¿qué unidades le corresponden a la constante $C$? ¿La dinámica lineal de partículas y cuerpos, es un caso particular de la dinámica curvilínea donde $\kappa = 0$?, si derivamos (19) respecto a las otras variables, ¿también encontraremos un punto de máxima curvatura? ¿la raíz imaginaria de (20), tiene algún sentido físico?, entre otras preguntas y casos particulares que se pueden formular a partir de la construcción teórica, sobre todo para analizar los detalles más sutiles en donde frecuentemente radica mucha física oculta. Desde luego que poder experimentar en un laboratorio de mecánica sería el complemento adecuado para demostrar la teoría, volviendo más cercanos los conceptos que se dedujeron geométricamente al inicio de este desarrollo.

## IV. CONCLUSIONES Y RECOMENDACIONES

Tal y como se mostró a lo largo del trabajo, el concepto de curvatura en un nivel de cálculo de los primeros semestres universitarios tiene la potencia de introducir conceptos de la física que el estudiante aprende de manera simultánea, lo cual brinda un mayor nivel de profundidad en el estudio de los temas matemáticos y físicos. Además, con ayuda de la transversalidad se estimula el análisis integral gracias al estudio de casos particulares, por ejemplo, los casos límites expuestos para la aceleración y fuerza centrípeta, y a la modelación gráfica del sistema físico, en la cual se pone de manifiesto la generalidad del sistema físico estudiado y facilita la interpretación y evaluación de los resultados analíticos y numéricos.

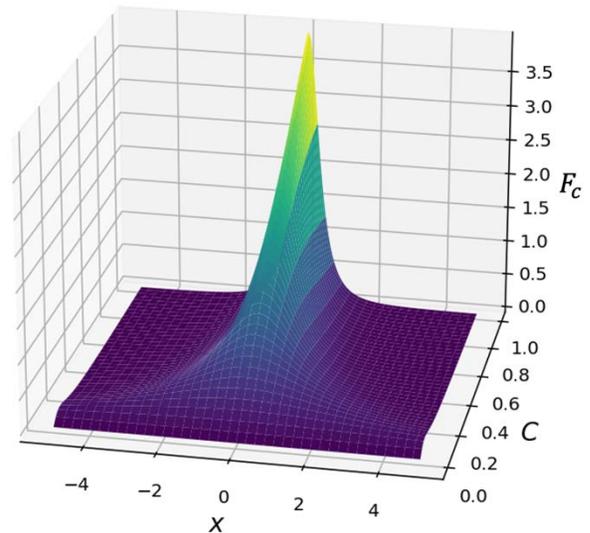

**FIGURA 8.** Comportamiento de la fuerza centrípeta en función de $x$ y $C$. Implícitamente se manifiesta la influencia de $\kappa$.

En la práctica docente, se recomienda utilizar conceptos clave y relativamente fácil de entender como lo es la definición geométrica de curvatura o aún más intuitivo, el radio de curvatura, para fortalecer o establecer un puente entre temas aparentemente inconexos, o cuya relación implique un dominio aún mayor, por ejemplo, de matemáticas más avanzadas. Lo anterior promoverá una mayor ilación en la clase teórica expuesta por el docente y además promoverá en el estudiante un pensamiento analítico y creativo sobre las diferentes formas de abordar un tema y las aplicaciones en propuestas de soluciones alternas.

Para que las secuencias didácticas transversales entre tópicos de matemáticas y física, o incluso de otros campos disciplinares tengan éxito, se debe fortalecer la relación interdepartamental de las instituciones, entre otras cosas, para buscar estrategias de intercambio de ideas, y de manera operativa en la docencia empatar temas para abordarlos de manera simultánea o casi simultánea, lo cual la mayoría de las veces es un reto logístico enorme para el docente, por lo que actualmente se sigue trabajando mayoritariamente en la práctica diaria de manera inconexa o con esfuerzos muy puntuales de los docentes.

Aprender las matemáticas de los últimos años de educación media y los primeros de la superior, dota al alumno de una herramienta para resolver problemas en física, lo cual parece ser el objetivo de muchos cursos (sólo resolver problemas), pero pensar en un sentido matemático promueve en el estudiante además de resolver ejercicios, a imaginar fuera de los límites o propósitos del problema; es decir, crear nuevas formas de abordarlo e interpretarlo. Tal y como vimos en este trabajo, se estudiaron temas clásicos de la mecánica con ayuda de un concepto matemático poco usual en esos cursos, e incluso se modeló el comportamiento de distintos casos particulares.

Se recomienda realizar evaluaciones comparadas de esta propuesta, midiendo específicamente la escala de aprendizaje de los conceptos matemáticos e interpretaciones físicas.

**AGRADECIMIENTOS**

**REFERENCIAS**